\documentclass[journal]{vgtc}                
\ifpdf
  \pdfoutput=1\relax                   
  \usepackage{graphicx}                
  \DeclareGraphicsExtensions{.pdf,.png,.jpg,.jpeg} 
\else
  \usepackage{graphicx}                
  \DeclareGraphicsExtensions{.eps}     
\fi%

\graphicspath{{figures/}{pictures/}{images/}{./}} 

\usepackage{microtype}                 
\PassOptionsToPackage{warn}{textcomp}  
\usepackage{textcomp}                  
\usepackage{mathptmx}                  
\usepackage{times}                     
\usepackage{cite}                      
\usepackage{tabu}                      
\usepackage{booktabs}                  

\usepackage{balance}  
\usepackage{graphics} 
\usepackage{times}    
\usepackage{url}      
\usepackage[ruled,vlined,linesnumbered]{algorithm2e}
\usepackage{enumitem}
\usepackage{xspace}
\usepackage[table,usenames,dvipsnames]{xcolor}
\usepackage{booktabs}
\usepackage{multirow}
\usepackage{array}
\usepackage{cellspace}
\usepackage[10pt]{moresize} 
\usepackage{amsmath}
\usepackage{listings}
\usepackage{xcolor}
\usepackage{amsmath}
\usepackage{MnSymbol}
\colorlet{punct}{red!60!black}
\definecolor{background}{HTML}{EEEEEE}
\definecolor{delim}{RGB}{20,105,176}
\colorlet{numb}{magenta!60!black}

\lstdefinelanguage{json}{
    basicstyle=\small\ttfamily,
    numberstyle=\scriptsize,
    showstringspaces=false,
    breaklines=true,
    frame=lines,
    literate=
     *{0}{{{\color{numb}0}}}{1}
      {1}{{{\color{numb}1}}}{1}
      {2}{{{\color{numb}2}}}{1}
      {3}{{{\color{numb}3}}}{1}
      {4}{{{\color{numb}4}}}{1}
      {5}{{{\color{numb}5}}}{1}
      {6}{{{\color{numb}6}}}{1}
      {7}{{{\color{numb}7}}}{1}
      {8}{{{\color{numb}8}}}{1}
      {9}{{{\color{numb}9}}}{1}
      {:}{{{\color{punct}{:}}}}{1}
      {,}{{{\color{punct}{,}}}}{1}
      {\{}{{{\color{delim}{\{}}}}{1}
      {\}}{{{\color{delim}{\}}}}}{1}
      {[}{{{\color{delim}{[}}}}{1}
      {]}{{{\color{delim}{]}}}}{1},
}

\onlineid{1150}

\vgtccategory{Research}
\vgtcpapertype{System}

\usepackage{ifthen}
\newboolean{revising}
\setboolean{revising}{false}
\ifthenelse{\boolean{revising}}
{
    \newcommand{\rv}[1]{\textcolor{blue}{#1}}
} {
    \newcommand{\rv}[1]{#1}
}

\newcommand{\name}{Calliope\xspace}
\newcommand{\etal}{{\it et~al.}\xspace}
\newcommand{\type}{\textit{\textbf{type}}\xspace}
\newcommand{\subspace}{\textit{\textbf{subspace}}\xspace}
\newcommand{\measure}{\textit{\textbf{measure}}\xspace}
\newcommand{\focus}{\textit{\textbf{focus}}\xspace}
\newcommand{\breakdown}{\textit{\textbf{breakdown}}\xspace}

\newcommand{\selection}{\textit{\textbf{selection}}\xspace}

\newcommand{\expansion}{\textit{\textbf{expansion}}\xspace}

\newcommand{\simulation}{\textit{\textbf{simulation}}\xspace}

\newcommand{\backprop}{\textit{\textbf{back-propagation}}\xspace}

\newcommand{\swiper}{\textit{\textbf{swiper}}\xspace}

\newcommand{\storyline}{\textit{\textbf{storyline}}\xspace}

\newcommand{\factsheet}{\textit{\textbf{factsheet}}\xspace}

\newcommand{\Covid}{\textit{\textbf{COVID-19}}\xspace}
\newcommand{\Startup}{\textit{\textbf{Startup Failures}}\xspace}
\newcommand{\CarSales}{\textit{\textbf{Car Sales}}\xspace}

\SetCommentSty{mycommfont}

\title{\name: Automatic Visual Data Story Generation \\ from a Spreadsheet}

\author{Danqing Shi, Xinyue Xu, Fuling Sun, Yang Shi and Nan Cao}
\authorfooter{
\item
 Danqing Shi, Xinyue Xu, Fuling Sun, Yang Shi, and Nan Cao are with Intelligent Big Data Visualization Lab at Tongji University.  
 
 E-mail: \{sdq, xuxinyue, fulingsun, yangshi.idvx\}@tongji.edu.cn, nan.cao@gmail.com. Nan Cao is the corresponding author.
}

\shortauthortitle{Biv \MakeLowercase{\textit{et al.}}: Global Illumination for Fun and Profit}

\abstract{%
Visual data stories shown in the form of narrative visualizations such as a poster or a data video, are frequently used in data-oriented storytelling to facilitate the understanding and memorization of the story content. Although useful, technique barriers, such as data analysis, visualization, and scripting, make the generation of a visual data story difficult. Existing authoring tools rely on users' skills and experiences, which are usually inefficient and still difficult. In this paper, we introduce a novel visual data story generating system, \name, which creates visual data stories from an input spreadsheet through an automatic process and facilities the easy revision of the generated story based on an online story editor. Particularly, \name incorporates a new logic-oriented Monte Carlo tree search algorithm that explores the data space given by the input spreadsheet to progressively generate story pieces (i.e., data facts) and organize them in a logical order. The importance of data facts is measured based on information theory, and each data fact is visualized in a chart and captioned by an automatically generated description. We evaluate the proposed technique through three example stories, two controlled experiments, and a series of interviews with 10 domain experts. Our evaluation shows that \name is beneficial to efficient visual data story generation. 
} 


\CCScatlist{ 
 \CCScat{K.6.1}{Management of Computing and Information Systems}%
{Project and People Management}{Life Cycle};
 \CCScat{K.7.m}{The Computing Profession}{Miscellaneous}{Ethics}
}

\teaser{
  \centering
  \includegraphics[width=\linewidth]{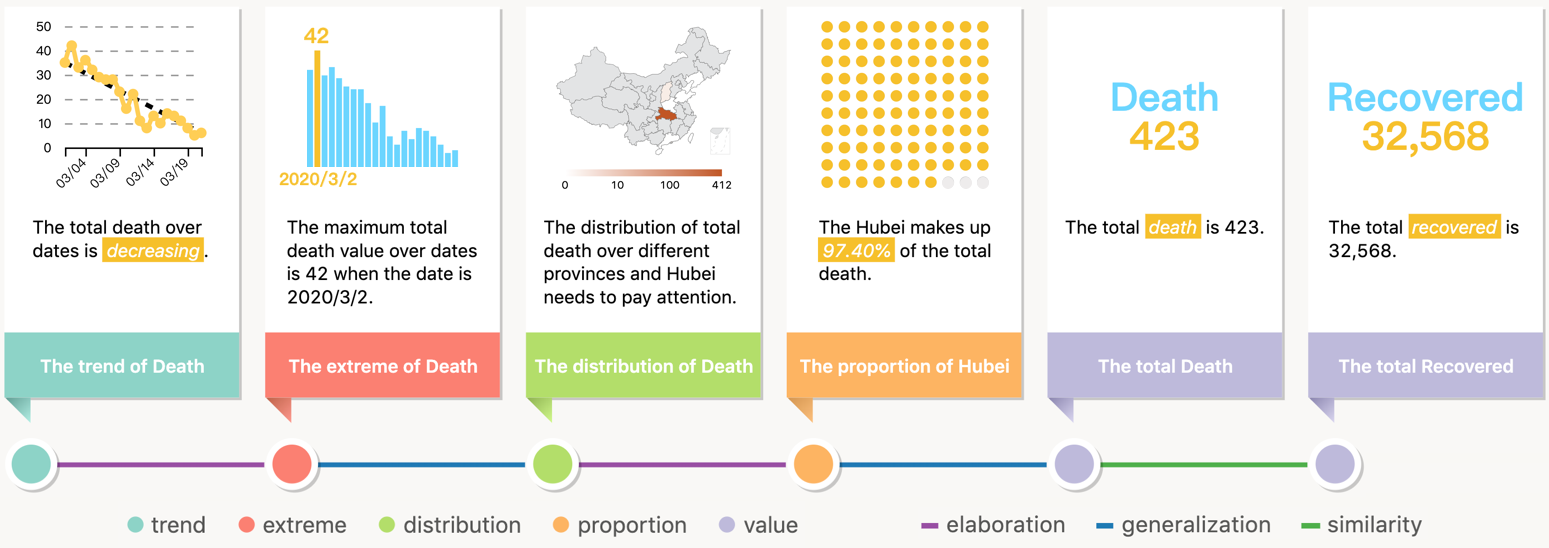}
  \caption{A full-automatically generated visual data story about COVID-19 virus propagation in March 2020 in China. The whole story consists of six data facts shown in the \storyline mode. The story first illustrates an overall trend of the total death followed by the elaborations of data details in Hubei, which is the most influenced province in China. Finally, the story is summarized by the total death and recovered in March, showing the improved situation in China.}
  \label{fig:teaser}
}

\vgtcinsertpkg

\begin{document}

\firstsection{Introduction}
\maketitle
Visual data story is a representation of a series of meaningfully connected story pieces (i.e., data facts) in the form of a narrative visualization~\cite{lee2015more,segel2010narrative}. Such form of representation is frequently used in a data-driven storytelling process due to its efficiency in terms of supporting the comprehension and memorization of the telling content~\cite{borkin2015beyond, hullman2013deeper}. Examples include various data news created by journalists in New York Times and Prof. Hans Rosling's talks on human development trends. Although useful, creating a visual data story requires an author to have a variety of skills, including data analysis, visualization, and scripting, which are usually difficult to perform for an ordinary user.

\rv{Technical barriers motivate the rapid development of a variety of authoring tools for creating various types of narrative visualizations, such as data videos~\cite{amini2016authoring,satyanarayan2019critical}, infographics~\cite{wang2018infonice}, and annotated charts~\cite{ren2017chartaccent}, to represent a data story. These tools usually incorporate built-in data analysis and visualization components to help with the story creation process. However, users must be involved to manually extract the data insights, write scripts, and assemble story pieces logically. Such involvement is usually inefficient and the quality of the resulting story still relies on the users' experiences and skills. Therefore, instead of developing another interactive authoring tool, a more intelligent system that can automatically generate a high-quality data story directly from the input data and support flexible story editing functions is desired by ordinary users who have limited data knowledge.}

However, designing such a technique is difficult given numerous challenges to be addressed. First, automatically finding a set of informative elementary story pieces for building up a data story is usually difficult. It not only requires a fast exploration in an enormous data space to create candidate facts, but also needs qualitative measurement to estimate the importance of each fact to make them comparable, enabling the selection of important story pieces. Second, a data story is usually presented by narrative visualizations accompanied with text narrations and annotations to precisely express the message and avoid ambiguity~\cite{lee2015more,segel2010narrative}. However, automatically generating human-understandable visual and textual representations for each data fact in a story and matching them to each other is a challenging task due to the complexity of the data and the automatic generation problem. Third, all the facts must be organized in a narratively logical order to make the story understandable and meaningful when generating a story. However, computing to generate a narrative logic based on structured data remains a difficult problem supported by limited previous research in the visualization field.

\rv{To address the above challenges, we introduce \name\footnote{\name is available online at \url{https://datacalliope.com}}, an intelligent system designed for automatically generating visual data stories from a spreadsheet.}
In the system, we introduce a logic-oriented Monte Carlo tree search (MCTS) algorithm to explore the space given by the input data and generate potential data facts for the story in a logical context while the exploration. The importance of the generated facts in each search step will be estimated by their information quantity, which is calculated based on the information theory~\cite{shannon1948a} and their pattern significance calculated based on auto-insight techniques~\cite{ding2019quickinsights, tang2017extracting, wang2019datashot}. To facilitate a fast revision on the generated story, \name provides an interactive story editor, through which a user can easily edit on the general story logic and the details of each data fact. The final story can be further published on cloud for communication and sharing. The major contributions of the paper are as follows:

\begin{itemize}[leftmargin=10pt,topsep=2pt]
\itemsep -.5mm
\item {\bf System.} We introduce the first system, to the best of our knowledge, that is designed to automatically generate visual data stories. The system also provides authoring and communication functions that enable easy story editing and sharing.

\item {\bf Story Generation Algorithm.} We introduce a logic-oriented Monte Carlo tree search algorithm that explores the data space to generate a series of data facts in logical context to build a story. The algorithm avoids the time-consuming enumeration of the data space via a reward function and a logic filter to ensure the quality of the generation results. The run-time of each searching step can also be precisely controlled to ensure efficiency.

\item {\bf Story Information Measurement.} We introduce the first method, to the best of our knowledge, which can precisely estimate the information quantity of a data story. Specifically, we calculate the self-information of each story fact and estimate the content of the story based on the information entropy by using and extending the definitions introduced in the information theory.

\item {\bf Evaluation.} We demonstrate the utility of the proposed system via three example stories generated based on real-world data, two controlled experiments designed to verify the generated logic, and a series of interviews with 10 expert users respectively from three areas, including data journalism, business intelligence, and the visualization research community. 
\end{itemize}

\section{Related Work}
\label{sec:related}
In this section, we review the recent studies that are most relevant to our work, including data-driven storytelling, automatic data visualization, and natural language generation.

\subsection{Data-Driven Storytelling} 
Data-driven storytelling is a rapidly developing research direction that focuses on techniques for enhancing data understanding, information expression, and communication. Narrative visualization is one promising approach frequently used for data-driven story telling~\cite{tong2018storytelling}. Recently, the visualization community has extensively investigated storytelling and narrative visualization techniques~\cite{segel2010narrative, tong2018storytelling}. According to Segel and Heer~\cite{segel2010narrative}, narrative visualization can be largely classified into seven genres, including magazine style, annotated chart, partitioned poster, flow chart, comic strip, slideshow, and video. Evidence shows that an effective composition and visual narrative of the story can guide readers through the data and improve the comprehension and memory~\cite{borkin2015beyond, hullman2013deeper}. 
To compose effective data-driven stories, Hullman~\etal~\cite{hullman2013deeper} identified several key design actions in the sequential story creation process, including context definition, facts selection, modality selection, and order selection.
Lee~\etal also decomposed the creation process of a data-driven story into three major steps: insights finding, scripting, and communicating to the audience~\cite{lee2015more}. These valuable study results guide the designs of many data story creation systems including \name, which is introduced in this paper.

Users commonly experience difficulty in creating a data-driven story due to technical barriers, which motivate the design and development of various authoring tools. General tools such as Ellipsis~\cite{satyanarayan2014authoring} allows a user to directly integrate visualizations to an illustrative story. Recent studies focus on designing tools to generate a specific type of visual narrative. For example, ChartAccent~\cite{ren2017chartaccent} and InfoNice~\cite{wang2018infonice} are respectively designed for creating annotated charts and infographics. Narvis~\cite{wang2018narvis} is introduced to extract the combination of visual elements of a visualization and organize them as a slideshow to help with the narrative interpretation of a visualization design. DataClips~\cite{amini2016authoring} is designed to help users generate data videos. Various authoring tools that are specifically designed to create narrative visualization for certain types of data. For example, Timeline Storyteller~\cite{brehmer2019timeline}, is a visual storytelling tool designed specifically for time-oriented data. Several visualization tools are also introduced to bridge the gap between visual analysis and storytelling~\cite{chen2018supporting, gratzl2016visual}, but they still target on expert users.

The above tools assume that the story content is manually created, resulting in inefficiency. By contrast, \name supports automatic story generation and flexible story editing functions, which ensure the quality, lower the barrier, and improve the efficiency of visual narration.

\subsection{Automatic Data Visualization}
Studies on automatic visualization have experienced three stages, including visualization chart recommendation, automatic data mapping generation, and auto-insights. To recommend a chart given an input data, early studies employed a rule-based method, which checks the data types to make a suggestion~\cite{mackinlay2007show,gotz2010harvest}. A recent study~\cite{hu2019vizml} trained a classification model based on a collection of ``data feature - chart type" pairs extracted from a visualization design corpus. As a result, given data features, a proper chart type can be selected. In \name, we select a chart for each data fact based on its fact type and data fields.

To visualize data in a chart, one must determine the detailed data mapping strategy. To this end, various techniques are introduced. Draco~\cite{moritz2018formalizing} uses an optimization model to find the best data mapping strategy under a set of constraints formulated by several common visual design guidelines. DeepEye~\cite{luo2018deepeye} enumerates all possible data mapping strategies and uses a decision tree to select the good ones. Data2Vis~\cite{dibia2019data2vis} ``translates" the data into a visual encoding scheme based on a sequence-to-sequence deep model. Text-to-Viz~\cite{cui2019text} employs natural language processing techniques to identify and parse data entities, such as numbers, portions, and data scopes from an input text, and convert them into a statistic diagram. Shi~\etal\cite{Shi2019TaskOrientedOS} explored the chart design space via a reinforcement learning model and generated a sequence of data mapping approaches regarding an analytical task. \name encodes different data fact fields in a chart following a rule-based method.

Recent studies focused on extracting data patterns and representing them in charts to reveal data insights, i.e., auto-insights~\cite{tang2017extracting,ding2019quickinsights}. The extracted insights can be quantitatively measured based on their statistical significance~\cite{ding2019quickinsights}. Visual analysis techniques were also developed to support auto-insights. For example, SeeDB~\cite{vartak2015seedb} finds and illustrates the most interesting trend in the data by exploring various data mapping strategies. Foresight~\cite{demiralp2017foresight} extracts and visualizes insights from multidimensional data using rule-based methods. DataShot~\cite{wang2019datashot} randomly visualizes a set of automatically generated data facts as a factsheet. \rv{Inspired by DataShot, \name also borrowed the auto-insights techniques to generate data facts, but made a step further by organizing the facts in a logical order to generate a meaningful data story.}

\subsection{Natural Language Generation}
Recently, studies on natural language generation (NLG) demonstrate the capability of producing descriptive text from various types of data~\cite{mishra2019storytelling,turner2009generating,galanis2007generating}. Many methods use templates to generate sentences~\cite{swanson2012say,li2013story}, and techniques that automatically enrich a template were developed~\cite{dou2018data2text,ye2020variational}. Recent studies leveraged deep learning models to generate textual content from scratch~\cite{fan2018hierarchical,fan2019strategies}. Several methods create an intermediate structure based on a recurrent neural network~\cite{martin2018event,xu2018skeleton,yao2019plan}, and others use the auto-encoder architecture to generate diverse sentences from a latent space~\cite{liu2019transformer,li2019learning}. Among various techniques, those aim to generate text content based on structured data are the most relevant to our work~\cite{mahapatra2016statistical,belz2008automatic,jain2018mixed}. For example, commercial software, such as PowerBI\footnote{\url{https://powerbi.microsoft.com}} and Quill\footnote{\url{https://narrativescience.com}} describe important data facts based on a set of templates to help interpret the data and the corresponding visualization. A number of visual auto-insights systems, such as DataSite~\cite{cui2019datasite}, DataShot~\cite{wang2019datashot}, and Voder~\cite{srinivasan2018augmenting}, use the template-based NLG to generate captions for visualization charts. In \name, we also employ the template-based method to generate captions for each chart, but to ensure the readability and avoid ambiguity, we define a syntax for each fact type that regulates the generation results.
\setlength{\textfloatsep}{20pt}
\begin{figure}[tb]
  \centering
  \includegraphics[width=\linewidth]{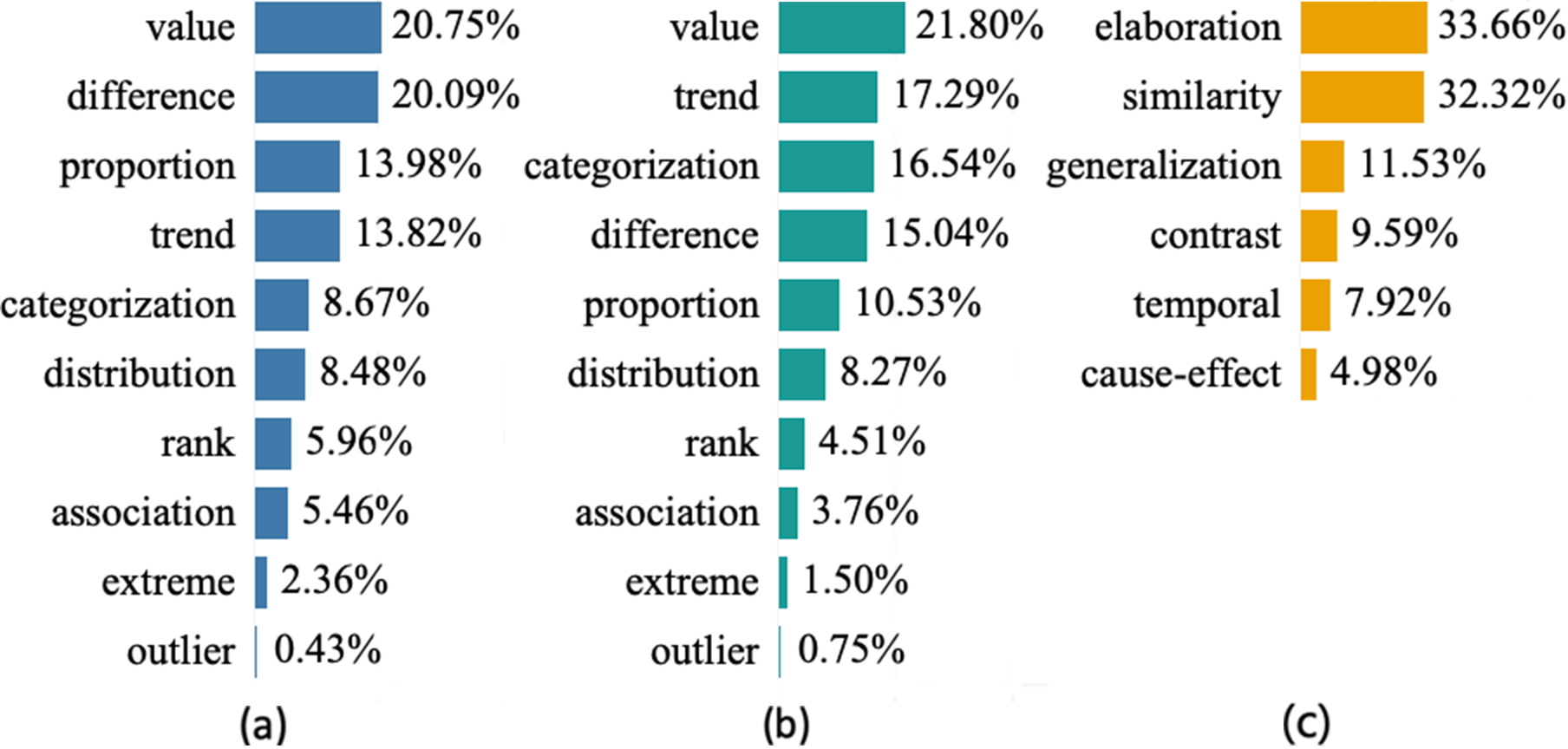}
  \vspace{-1.5em}
  \caption{The statistics on the labeled data video corpus showing the percentage of (a) data facts, (b) fact types used as the start points, and (c) narrative relationships.} 
  \label{fig:statistic}
  \vspace{+0.5em}
\end{figure}
\setlength{\textfloatsep}{20pt}
\section{Design of the Calliope System}
\label{sec:system}
This section introduces the design of \name system. We first introduce the formal definition of a data story, then survey on a collection of data videos to help us understand how a story is generated by human designers. After that, we summarize the design requirements and introduce the architecture design of the system.

\vspace{-0.5em}
\begin{table}[tb]
\setlength\aboverulesep{0pt}
\setlength\belowrulesep{0pt}
\centering
\small
\def\arraystretch{1.5}
\setlength\leftskip{0pt plus 1 fil minus \marginparwidth}
\makebox[\linewidth][c]{
\begin{tabular}{cwc{0.1cm}wc{0.1cm}wc{0.1cm}wc{0.1cm}wc{0.1cm}wc{0.1cm}wc{0.1cm}wc{0.1cm}wc{0.1cm}wc{0.1cm}wc{0.1cm}wc{0.1cm}}
\toprule
\rotatebox{0}{Fact Types}      & \rotatebox{60}{Bar} & \rotatebox{60}{Line} & \rotatebox{60}{ISOType} & \rotatebox{60}{Map} & \rotatebox{60}{Scatter plot} & \rotatebox{60}{Pie} & \rotatebox{60}{Area} & \rotatebox{60}{Bubble} & \rotatebox{60}{Text} & \rotatebox{60}{Table} & \rotatebox{60}{Box plot} & \rotatebox{60}{Treemap} \\\midrule

Value          &  \cellcolor[HTML]{99d8c9}56  &      &  \cellcolor[HTML]{ccece6}26      &  \cellcolor[HTML]{99d8c9}40  &             &     &      &        &  \cellcolor[HTML]{238b45}204    &  \cellcolor[HTML]{e5f5f9}9     &          &         \\
Difference     & \cellcolor[HTML]{41ae76}99  & \cellcolor[HTML]{ccece6}37   & \cellcolor[HTML]{ccece6}23      & \cellcolor[HTML]{ccece6}39  &             &     &      &        & \cellcolor[HTML]{99d8c9}46     &       &          &         \\
Proportion     & \cellcolor[HTML]{99d8c9}45  &      & \cellcolor[HTML]{66c2a4}68      & \cellcolor[HTML]{e5f5f9}9   &             & \cellcolor[HTML]{238b45}131 &      &        & \cellcolor[HTML]{99d8c9}58     &       &          &         \\
Trend          & \cellcolor[HTML]{66c2a4}60  & \cellcolor[HTML]{238b45}170  & \cellcolor[HTML]{e5f5f9}12      & \cellcolor[HTML]{e5f5f9}17  &             &     & \cellcolor[HTML]{e5f5f9}14   &        &        &       &          &         \\
Categorization & \cellcolor[HTML]{ccece6}24  &      & \cellcolor[HTML]{e5f5f9}4       & \cellcolor[HTML]{e5f5f9}12  &             &     & \cellcolor[HTML]{e5f5f9}4    &        &        &       &          & \cellcolor[HTML]{e5f5f9}5       \\
Distribution   & \cellcolor[HTML]{99d8c9}40  &      & \cellcolor[HTML]{ccece6}28      & \cellcolor[HTML]{238b45}107 &             &     & \cellcolor[HTML]{e5f5f9}12   &        &        &       &          &         \\
Rank           & \cellcolor[HTML]{ccece6}24  &      &         & \cellcolor[HTML]{e5f5f9}15  &             & \cellcolor[HTML]{e5f5f9}3   &      &        & \cellcolor[HTML]{ccece6}22     & \cellcolor[HTML]{e5f5f9}5     &          &         \\
Association    &     & \cellcolor[HTML]{e5f5f9}17   &         & \cellcolor[HTML]{e5f5f9}11  & \cellcolor[HTML]{e5f5f9}9           &     &      & \cellcolor[HTML]{e5f5f9}2      & \cellcolor[HTML]{e5f5f9}1      &       &          &         \\
Extreme        & \cellcolor[HTML]{e5f5f9}12  & \cellcolor[HTML]{e5f5f9}2    & \cellcolor[HTML]{e5f5f9}4       & \cellcolor[HTML]{e5f5f9}10  &             &     &      &        & \cellcolor[HTML]{e5f5f9}4      &       &          &         \\
Outlier        &     &      & \cellcolor[HTML]{e5f5f9}1       & \cellcolor[HTML]{e5f5f9}1   &             &     & \cellcolor[HTML]{e5f5f9}4    &        & \cellcolor[HTML]{e5f5f9}2      &       & \cellcolor[HTML]{e5f5f9}2        &
\\\bottomrule
\end{tabular}}
\vspace{0.5em}
\caption{The frequency of visualization charts regarding the 10 fact types.}
\label{tab:charts}
\end{table}
\setlength{\floatsep}{5pt}

\begin{table}[tb]
\setlength\aboverulesep{0pt}
\setlength\belowrulesep{0pt}
\renewcommand{\arraystretch}{0.2}
\centering
\small
\def\arraystretch{1.5}
\setlength\leftskip{0pt plus 1 fil minus \marginparwidth}
\begin{tabular}{ccccccc}
\toprule
\rotatebox{0}{Fact Types}         & \rotatebox{0}{$r_s$} & \rotatebox{0}{$r_t$} & \rotatebox{0}{$r_c$} & \rotatebox{0}{$r_a$} & \rotatebox{0}{$r_e$} & \rotatebox{0}{$r_g$} \\\midrule
Value          & \cellcolor[HTML]{4292C6}45.6 & \cellcolor[HTML]{EFF3FF}8.9  & \cellcolor[HTML]{FFFFFF}0.0  & \cellcolor[HTML]{EFF3FF}4.2  & \cellcolor[HTML]{9ECAE1}26.8 & \cellcolor[HTML]{C6DBEF}14.5 \\
Difference     & \cellcolor[HTML]{4292C6}41.6 & \cellcolor[HTML]{EFF3FF}6.7  & \cellcolor[HTML]{FFFFFF}0.0  & \cellcolor[HTML]{EFF3FF}5.8  & \cellcolor[HTML]{6BAED6}31.1 & \cellcolor[HTML]{C6DBEF}14.8 \\
Proportion     & \cellcolor[HTML]{4292C6}52.1 & \cellcolor[HTML]{EFF3FF}7.3  & \cellcolor[HTML]{FFFFFF}0.0  & \cellcolor[HTML]{EFF3FF}5.2  & \cellcolor[HTML]{9ECAE1}22.4 & \cellcolor[HTML]{C6DBEF}13.0 \\
Trend          & \cellcolor[HTML]{6BAED6}34.7 & \cellcolor[HTML]{EFF3FF}9.4  & \cellcolor[HTML]{EFF3FF}8.2  & \cellcolor[HTML]{EFF3FF}7.1  & \cellcolor[HTML]{9ECAE1}28.2 & \cellcolor[HTML]{C6DBEF}12.4 \\
Categorization & \cellcolor[HTML]{9ECAE1}37.7 & \cellcolor[HTML]{EFF3FF}3.4  & \cellcolor[HTML]{FFFFFF}0.0  & \cellcolor[HTML]{EFF3FF}3.4  & \cellcolor[HTML]{4292C6}47.5 & \cellcolor[HTML]{EFF3FF}7.8  \\
Distribution   & \cellcolor[HTML]{4292C6}49.0 & \cellcolor[HTML]{C6DBEF}12.1 & \cellcolor[HTML]{FFFFFF}0.0  & \cellcolor[HTML]{EFF3FF}4.4  & \cellcolor[HTML]{9ECAE1}22.3 & \cellcolor[HTML]{C6DBEF}12.1 \\
Rank           & \cellcolor[HTML]{4292C6}43.8 & \cellcolor[HTML]{C6DBEF}11.7 & \cellcolor[HTML]{FFFFFF}0.0 & \cellcolor[HTML]{EFF3FF}6.6  & \cellcolor[HTML]{6BAED6}34.3 & \cellcolor[HTML]{EFF3FF}3.6  \\
Association    & \cellcolor[HTML]{6BAED6}31.0 & \cellcolor[HTML]{EFF3FF}5.6  & \cellcolor[HTML]{C6DBEF}15.1 & \cellcolor[HTML]{EFF3FF}7.1  & \cellcolor[HTML]{9ECAE1}26.2 & \cellcolor[HTML]{C6DBEF}15.1 \\
Extreme        & \cellcolor[HTML]{4292C6}51.8 & \cellcolor[HTML]{EFF3FF}5.6  & \cellcolor[HTML]{FFFFFF}0.0 & \cellcolor[HTML]{EFF3FF}3.7  & \cellcolor[HTML]{9ECAE1}25.9 & \cellcolor[HTML]{C6DBEF}13.0 \\
Outlier        & \cellcolor[HTML]{9ECAE1}20.0 & \cellcolor[HTML]{C6DBEF}10.0 & \cellcolor[HTML]{FFFFFF}0.0  & \cellcolor[HTML]{C6DBEF}10.0 & \cellcolor[HTML]{4292C6}40.0 & \cellcolor[HTML]{9ECAE1}20.0 
\\\bottomrule
\end{tabular}
\vspace{0.5em}
\caption{The likelihood of coherence relations, including similarity ($r_s$), temporal ($r_t$), contrast ($r_c$), cause-effect ($r_a$), elaboration ($r_e$), and generalization ($r_g$) used after the 10 fact types.}
\label{tab:logic}
\vspace{-1em}
\end{table}
\setlength{\textfloatsep}{5pt}

\subsection{Data Story}
Data story is a set of story pieces that are meaningfully connected to support the author's communication goal~\cite{lee2015more}. A story piece is a fact backed up by data, and it is usually visualized by succinct but expressive charts, accompanied with annotations (labels, pointers, text, etc.) and narrations to express the message and avoid the ambiguity. We design \name to automatically generate visual data stories by following this definition. Formally, a data story $\mathcal{S}$ consists of a sequence of data facts that are connected by coherent relations (denoted as $r_i \in \mathcal{R}$) $\{f_1,r_1, \cdots, f_{n-1}, r_{n-1}, f_n\}$ with each fact $f_i \in F$. We will use these notations throughout the paper. 

\subsection{Preliminary Survey}
Before designing the system, it is necessary to understand how a data story is created by a human designer. To this end, as data video is a frequently used narrative visualization form, we collected a set of 602 data videos from YouTube and Vimeo by searching keywords, such as ``animated infographic'', ``data video'', and ``motion infographic''. A total of 230 high-quality videos were selected and manually segmented into 4186 story pieces. The fact and chart types of 2583 data-related story pieces and the coherence relations used for connecting two succeeding pieces were labeled for analysis. \rv{Here, we borrowed the definition of fact types introduced in DataShot~\cite{wang2019datashot} and coherence relations introduced in~\cite{wellner2006classification, wolf2005representing} to label our data}. 

As a result, even some simple statistics are able to help us answer a number of questions that are critical to the design of an automatic story generation system. For example, Fig.~\ref{fig:statistic} suggests the frequently used fact types, the fact types frequently used as start points, and the frequently used coherence relations regarding our data video corpus. Table~\ref{tab:charts} suggests the preferred visualization charts given a fact type. Table~\ref{tab:logic} suggests the narrative logic probably used following each fact type. These results guide the design of the story generation algorithm and will be further described later. 

\subsection{System Design}
Our goal is to design a system that can automatically generate high-quality initial data stories directly from an input spreadsheet and support flexible story editing functions to lower the technical barriers of creating a data story. To achieve this goal, a number of key requirements need to be fulfilled:

\begin{enumerate}
\itemsep -1mm
\item[{\bf R1}] {\bf Generating ``successful" stories.} The most important thing for the system is to ensure the quality of story generation. Among various factors that contribute to a successful narrative artifact, the key is understandability~\cite{riedl2010narrative}, which is usually determined by the narrative logic, and the meaningful and believable content~\cite{lee2015more}. Therefore, the system should be intelligent enough to automatically generate meaningful stories logically with correct, i.e., believable, data backups.

\item[{\bf R2}] {\bf Efficient story generation.} The system should be able to efficiently generate a data story within a reasonable period of time that is affordable to the users. Therefore, the generation time should be controllable to grantee the efficiency of the system while keeping the quality of the story.

\item[{\bf R3}] {\bf Expressive story representation.} As suggested in~\cite{lee2015more}, the generated visual data story should be expressively represented in both visual and textual forms to precisely express the message and avoid ambiguity. Here, simple but intuitive charts~\cite{wang2019datashot}, as well as precise and meaningful narratives~\cite{lee2015more} should be guaranteed to reduce users' learning efforts.

\item[{\bf R4}] {\bf Easy story editing.} The system should provide flexible interactions to support comprehensive editing of the generated storyline, text narration, visual representation, and the corresponding data facts, so that a user can further refine and adjust an automatically generated story based on their own requirements.

\item[{\bf R5}] {\bf Easy communication and sharing.} The visual and textual representations of a data story should be probably aligned and adaptive laid out to fit into different devices such as a laptop, tablets, and smartphones to facilitate an easy story exploration, communication, and sharing.
\end{enumerate}

\begin{figure}[tb]
  \centering
  \includegraphics[width=\linewidth]{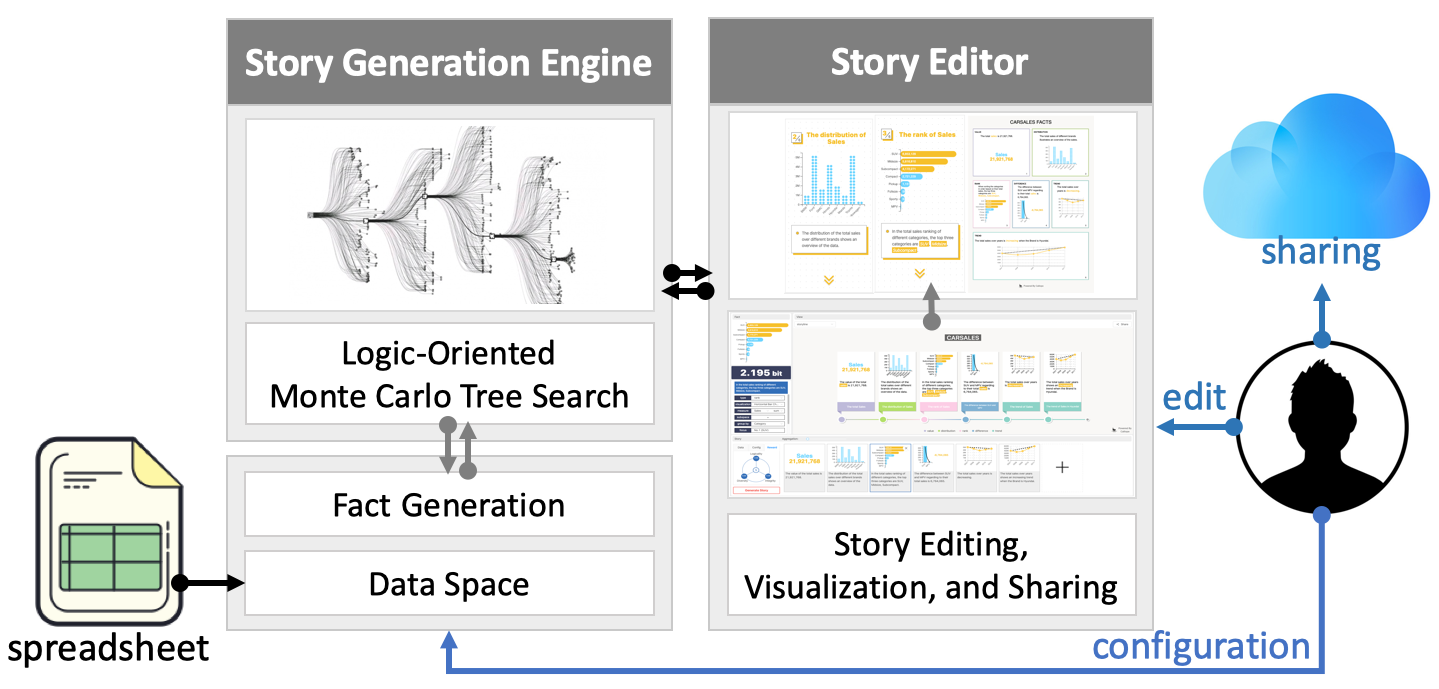}
  \vspace{-2em}
  \caption{\name system consists of two modules : the story generation engine and the story editor.} \label{fig:system}
  \vspace{-1.5em}
\end{figure}

To fulfill these requirements, the design of \name system consists of two modules (Fig.~\ref{fig:system}): (1) the story generation engine and (2) the story editor. The story generation engine is designed based on a logic-oriented Monte Carlo tree search process, in which a story is gradually generated fact by fact while searching through the data space defined by an input spreadsheet. The whole search process is guided by narrative logic and a reward function that measures the importance of facts to ensure the quality of the generated story (\textbf{R1}). In addition, the time spent on each searching step is configurable, which guarantees the generation efficiency (\textbf{R2}). The generated story is visualized in the story editor as a series of captioned visualization charts (\textbf{R3}), whose data facts, caption, chart type, and logic orders can be revised according to user preferences (\textbf{R4}). The final visual data story can be represented in three modes to fit different devices (\textbf{R5}).

\setlength{\textfloatsep}{20pt}
\setlength{\textfloatsep}{20pt}
\section{Story Generation Engine}
\label{sec:engine}
In this section, we first formally define a data fact and its importance measurement. We then describe the details of the proposed automatic story generation algorithm.

\begin{table*}[t]
    \small
    \setlength\aboverulesep{0pt}
    \setlength\belowrulesep{0pt}
    \def\arraystretch{1.0}
\begin{tabular}{|cwc{1.35cm}wc{1.35cm}wc{1.35cm}wc{1.35cm}|wc{2.06cm}|p{6.2cm}|}
\toprule
\textbf{Fact Type} ($t_i$)    & \textbf{Subspace}  ($s_i$)    & \textbf{Breakdown}  ($b_i$)   & \textbf{Measure} ($m_i$)  & \textbf{Focus} ($x_i$)  & \textbf{Derived Value ($V_d$)}    & \textbf{Syntax}                                                                                                                                                    \\\midrule
Value & * & $\bigtimes$ & $N$   & $\bigtimes$ & derived value & The \{\{agg\}\} \{\{$m_i$\}\} is \{\{$V_d$\}\} when \{\{$s_i$\}\}.                                        \\\hline

Difference   & *  & C/T & $N$  & =2       & difference value
& \begin{tabular}[c]{@{}l@{}}The difference between \{\{$x_i${[}1{]}\}\} and \{\{$x_i${[}2{]}\}\} regarding to  \\ their \{\{agg\}\} \{\{$m_i$\}\} is \{\{$V_d$\}\} when \{\{$s_i$\}\}.\end{tabular}                    \\\hline

Proportion  & *  & C/T  & $N$  & =1  & percentage   & \begin{tabular}[c]{@{}l@{}}The \{\{$x_i$\}\} accounts for \{\{$V_d$\}\} of the \{\{agg\}\} \{\{$m_i$\}\} \\ when \{\{$s_i$\}\}.\end{tabular}                            \\\hline

Trend     & *    & T  & $N$   & $\geq 0$    &increasing/decreasing       & \begin{tabular}[c]{@{}l@{}}The \{\{$V_d$\}\} trend of \{\{agg\}\} \{\{$m_i$\}\} over \{\{$b_i$\}\}(s) when \\ \{\{$s_i$\}\} and the values of \{\{$x_i$\}\} needs to pay attention.\end{tabular}  \\\hline

Categorization & * & C & $\bigtimes$    & $\geq 0$  & number of categories       & \begin{tabular}[c]{@{}l@{}}There are \{\{$V_d$\}\} \{\{$b_i$\}\}(s) which are \{\{$C_1$\}\}, \{\{$C_2$\}\}, \\ \{\{...\}\}, and \{\{$C_n$\}\} when \{\{$s_i$\}\}, among which \{\{$x_i$\}\} \\ needs to pay attention.\end{tabular}  \\\hline

Distribution & * & C & $N$  & $\geq 0$    & $\bigtimes$ & \begin{tabular}[c]{@{}l@{}}The distribution of the \{\{agg\}\} \{\{$m_i$\}\} over \{\{$b_i$\}\}(s) when \\  \{\{$s_i$\}\} and \{\{$x_i$\}\} needs to pay attention.\end{tabular}
\\\hline

Rank     & *    & C/T   & $N$   & =3 (top 3) &  $\bigtimes$
& \begin{tabular}[c]{@{}l@{}}In the \{\{agg\}\} \{\{$m_i$\}\} ranking of different \{\{$b_i$\}\}(s), \\ the top three \{\{$b_i$\}\}(s) are \{\{$x_i${[}0{]}\}\}, \{\{$x_i${[}1{]}\}\}, \{\{$x_i${[}2{]}\}\},\\  when \{\{$s_i$\}\}.\end{tabular} 
\\\hline

Association  & *  & C/T & $N \times N$      &  $\bigtimes$    & correlation coefficient    & \begin{tabular}[c]{@{}l@{}}The Pearson correlation between the \{\{$m_i${[}1{]}\}\} and the \\ \{\{$m_i${[}2{]}\}\} is \{\{$V_d$\}\} when \{\{$s_i$\}\}.\end{tabular}
\\\hline

Extreme    & *  & C/T  & $N$     & =1 (max/min) 
& maximum/minimum   & \begin{tabular}[c]{@{}l@{}}The \{\{$V_d$\}\} value of the \{\{agg\}\} \{\{$m_i$\}\} is \{\{$x_i$\}\} when \\  \{\{$s_i$\}\}.\end{tabular}.
\\\hline

Outlier & * & C/T  & $N$   & =1 (outlier) 
& outlier score & \begin{tabular}[c]{@{}l@{}}The\{\{agg\}\} \{\{$m_i$\}\} of \{\{$x_i$\}\} is an outlier when compare \\ with that of other \{\{$b_i$\}\}(s) when \{\{$s_i$\}\}.\end{tabular}                                          \\\bottomrule          
\end{tabular}
\vspace{0.5em}
    \caption{Definitions of the 10 types of data facts, where $N,C,T$ respectively indicate the numerical, categorical, and temporal data types. \{agg\} is one of the following terms that indicates an aggregation method: total number of (i.e., count), total value of (i.e., sum), average, maximum, and minimum.}
    \label{tab:fact}
 \vspace{-2em}
\end{table*}

\subsection{Data Facts}
\label{sec:fact}
Data facts are the elementary building blocks of a data story. Each of these facts represents a piece of information extracted from the data. \rv{We first give a formal definition of data facts by simplifying the concepts introduced in~\cite{wang2019datashot,chen2009toward} to guarantee a clear semantic and then introduce a novel method used to estimate the importance of a given data fact.}

\paragraph{\bf Definition} A data fact is designed to measure a collection of data items in a subspace of an input dataset based on a measurable data field. The data items can be further divided into groups via a breakdown method. Formally, a fact, $f_i \in F$, is defined by a 5-tuple:
\[
\begin{aligned}
    f_i &= \{type, subspace, breakdown, measure, focus\}\\
    &= \{t_i, s_i, b_i, m_i, x_i\}
\end{aligned}
\]
\rv{where \type (denoted as $t_i$) indicates the type of information described by the fact. As summarized in Table~\ref{tab:fact}, \name includes 10 fact types;}
\subspace (denoted as $s_i$) describes the data scope of the fact, which is defined by a set of data filters in the following form: 
\[
\{\{\mathcal{F}_1 = \mathcal{V}_1\},\cdots, \{\mathcal{F}_k = \mathcal{V}_k\}\}
\]
where $\mathcal{F}_i$ and $\mathcal{V}_i$ respectively indicate a data field and its corresponding value selected to filter the data. By default, the subspace is the entire dataset. \breakdown (denote as $b_i$) is a set of temporal or categorical data fields based on which the data items in the subspace are further divided into groups; \measure (denote as $m_i$) is a numerical data field based on which we can retrieve a data value or calculate a derived value, such as count, sum, average, minimum, or maximum, by aggregating the subspace or each data group; \focus (denote as $x_i$) indicates a set of specific data items in the subspace that require attention. Despite the above five fields, certain facts may also have a \textit{\textbf{derived value}} (denoted as $V_d$) such as a textual summary of the trend (i.e., ``increasing" or ``decreasing") or the specific difference value between two cases described by a difference fact, or the correlation coefficient computed for an association fact as shown in Table~\ref{tab:fact}. These values help with a more insightful description of the fact.

\rv{When compared to the concepts introduced in~\cite{wang2019datashot,chen2009toward}, the above definition simplified and restricted the fact fields to ensure a clear semantic expression of the data that avoids redundancy, overlaps, and ambiguity. Specifically, when compared to \cite{wang2019datashot}, we removed the fact fields that are irrelevant to the fact semantics  and treated ``aggregation" as an operation on ``measures" instead of a fact type to avoid duplicated fact definitions. In addition, as summarized in Table~\ref{tab:fact}, we add constraints on each fact field to ensure a clear semantics. For example, the facts in ``distribution" and ``trend" types are designed to capture the data patterns given by the measures of different data groups in the subspace. Both fact types can be differentiated by their ways of breaking down a subspace: the subspace in a ``trend" fact must be divided by a temporal field, whereas the subspace in a ``distribution" fact can only be divided by a categorical field. Thus, each fact can be described by a syntax which is used for generating a textual description of the fact.}

To understand the above concepts, let's consider the following examples. Given a dataset about the COVID-19 virus outbreak in China, the data fact, {\it \{``distribution", \{\{Country =``China"\}\}, \{Province\}, \{sum(Infections)\}, \{Province=``Hubei"\}\}}, describes ``the distribution of the \underline{total number of} \underline{infections} over all \underline{provinces} when \underline{the country is China} (subspace) and \underline{Hubei} needs to pay attention" regarding to the syntax of the distribution fact. Similarly, the data fact, {\it \{``trend", \{\{Province =``Hubei"\}\}, \{Date\}, \{sum(Infections)\}, \{Date=``2020-1-24"\}\}}, indicates ``the changing trend of the \underline{total number of} \underline{infections} over different \underline{dates} when \underline{province is Hubei} and the values of \underline{2020-1-24} need to pay attention".



\paragraph{\bf Importance Score} We estimate the importance of a data fact $f_i = \{t_i,s_i,b_i,m_i,x_i\} \in F$ based on its \textit{self-information} (denoted as $I(f_i)$) weighted by its \textit{pattern significance} (i.e., $S(f_i)\in [0,1]$) as follows:
\begin{equation}
    I_s(f_i) = S(f_i) \cdot I(f_i)
\label{eq:importance}
\end{equation}

In particular, $I(f_i) \in [0, \infty]$ is defined based on the information theory and can be measured in ``bit" using the following formula:
\begin{equation}
I(f_i) = -log_2(P(f_i))
\label{eq:info}
\end{equation}
\rv{where $P(f_i)$ indicates the occurrence probability of the fact given the input data. A data fact with a lower occurrence probability in the data space has a higher self-information value as it reveals uncommon patterns which are usually more meaningful and interesting.} $P(f_i)$ is formally determined by the occurrence probability of the fact's  \subspace $s_i$, the probabilities of selecting $x_i$ as the \focus in $s_i$, and the probabilities of choosing $m_i$ ($P(m_i|t_i)$) and $b_i$ ($P(b_i|t_i)$) to measure and break down $s_i$ given a fact type $t_i$: 
\begin{equation}
P(f_i) = P(m_i|t_i) \cdot P(b_i|t_i) \cdot P(s_i) \cdot P(x_i | s_i)
\label{eq:pfi}
\end{equation}
where $P(m_i|t_i)$ and $P(b_i|t_i)$ is defined regarding the data type constraints of $m_i$ and $b_i$ as summarized in Table~\ref{tab:fact}. For example, when the fact type is ``Value", $P(m_i|Value)$ is $1/N$, where $N$ is the total number of numerical fields in the data. Similarly, $P(b_i|Difference)$ is $1/(C + T)$ where $C,T$ are the total number of categorical and temporal fields in the data. Moreover, in Formula (\ref{eq:pfi}), $P(x_i | s_i)$ is defined as the proportion of the focused data items in the subspace, i.e., $P(x_i | s_i) = count(x_i) / count(s_i)$, with the assumption that the probability of selecting each data item as a focus in the subspace is equivalent. In our design, all the data items in $s_i$ are focused by default, i.e., $P(x_i | s_i) = 1$ when the \focus field is unspecified. To calculate $P(s_i)$, we first assume $s_i$ consists of $k$ data filters, i.e., $\{\{\mathcal{F}_1 = \mathcal{V}_1\},\cdots, \{\mathcal{F}_k = \mathcal{V}_k\}\}$ and there are $m$ independent data fields that can be used for formulating a subspace. In this way, $P(s_i)$ is defined as follows:
\begin{equation}
P(s_i) = \frac{1}{\sum_{i=0}^{m} C(m, i)}\cdot\prod_{j=1}^{k}P(\mathcal{F}_j = \mathcal{V}_j)
\label{eq:p_si}
\end{equation}
where the first term indicates the probability of choosing the fields $\mathcal{F}_1,\cdots, \mathcal{F}_k$ from the input data to formulate the subspace $s_i$. $C(m, i)$ is an $i$-combination over a set of $m$ possible data fields. $\sum_{i=0}^{m} C(m, i)$ summarizes the number of all possible cases for formulating a subspace. In this way, the first term in Formula (\ref{eq:p_si}) indicates the method we use for formulating the current subspace $s_i$ is just one possible case. The second term in Formula (\ref{eq:p_si}) indicates the probability of using the corresponding values $\mathcal{V}_1,\cdots, \mathcal{V}_k$ on the selected fields to filter the data. This probability is directly given by the products of the proportions of the data that satisfy each filter conditions, i.e.,  $\{\mathcal{F}_j = \mathcal{V}_j\}$.

In Formula (\ref{eq:importance}), $S(f_i) \in [0, 1]$ estimates the significance of the data patterns described by the fact $f_i$, which is calculated based on auto-insight techniques~\cite{ding2019quickinsights, tang2017extracting, wang2019datashot}. The detailed methods are described in the supplemental material. \rv{It worth mentioning that a significant pattern may not necessary have a high self-information value. Only using both measurements as shown in Formula (\ref{eq:importance}) will guarantee a comprehensive estimation. Under this definition, the importance of a fact is only determined by its data content and irrelevant to how frequently a type of fact is used in data stories (Fig.~\ref{fig:statistic}(a)).}


\subsection{Story Generation Algorithm}
In \name, we introduce an intelligent algorithm that generates data facts from an input spreadsheet and threads them logically to create a meaningful data story. However designing such an algorithm is challenging. The story design space, formulated by a collection of data facts generated from the input spreadsheet, could be extremely large due to the huge number of possible combinations of the fact fields even based on a small dataset. The algorithm cannot generate all facts first and then pick up those important ones to build the story as users cannot afford a long waiting time. We addressed this issue by introducing an efficient logic-oriented searching algorithm based on the Monte Carlo tree search (MCTS)~\cite{browne2012survey,silver2016mastering}. The algorithm can efficiently explore a large space that contains a huge number of states via a searching process organized by a tree structure and guided by a reward function towards a logic direction.

\begin{figure}[!b]
  \centering
  \vspace{-0.5em}
  \includegraphics[width=\linewidth]{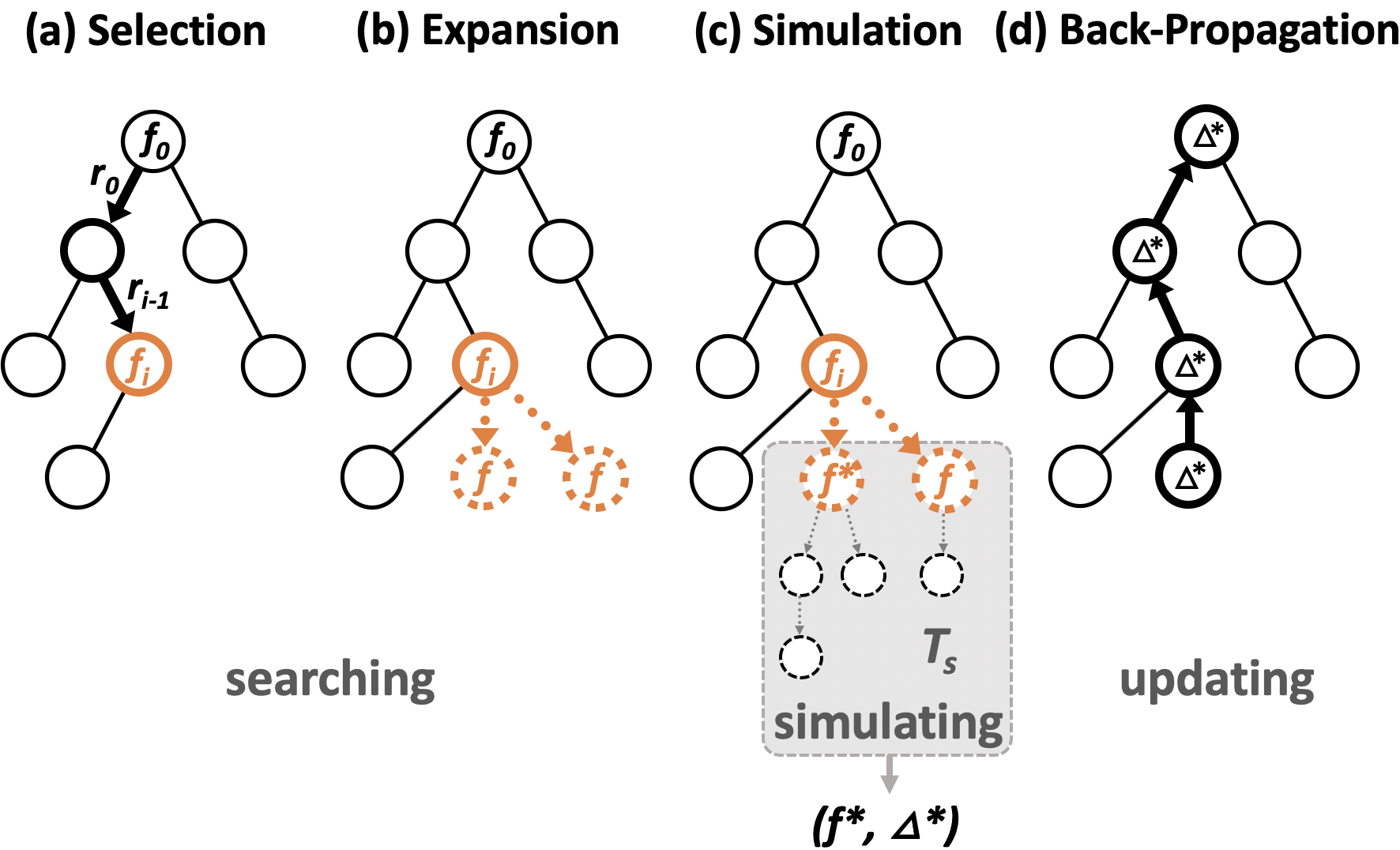}
  \vspace{-1em}
  \caption{An iteration of the logic-oriented Monte Carlo tree search consists of four steps, including (a) selection, (b) expansion, (c) simulation, and (d) back-propagation.} 
  \label{fig:algorithm}
  \vspace{-0.5em}
\end{figure}

\paragraph{\bf Algorithm Overview} In general, the algorithm explores the design space by dynamically constructing a searching tree $\mathcal{T}$. As shown in Fig.~\ref{fig:algorithm}(a), each node in the tree is a data fact $f_i$ and each directed edge indicates a logic relation $r_i$.  A data story $\mathcal{S} = \{f_1, r_1, \cdots,f_{n-1}, r_{n-1}, f_n\}$ is thus represented by a path starting from the root. A reward function is designed to estimate the quality of each path in the tree. The reward scores are marked on the last node in paths. A node shared by multiple paths is weighted by the maximum reward. These scores are used to guide the exploration of the design space.


The tree $\mathcal{T}$ is gradually generated through a searching process as described in Algorithm~\ref{alg:generation}. In particular, the algorithm takes a spreadsheet $\mathcal{D}$, i.e., a data table, and a goal $\mathcal{G}$, such as generating a story with a desired information quantity or length as the inputs and automatically generates a story $\mathcal{S}$ that fulfills the goal. \rv{Initially, it randomly generates a set of facts in types that are frequently used as the starting point in a data story (Fig.~\ref{fig:statistic}(b)). These facts usually reveal general and common data patterns which may already known by the audience as the background of the story. Among these facts, the most important one, denoted as $f_0$, is used as the root of $\mathcal{T}$. In the next, the algorithm generates a story by iteratively searching more informative and significant data facts to elaborate the story via four major steps: \selection, \expansion, \simulation, and \backprop.}
The first step finds a node $f_i$ with the largest reward in $\mathcal{T}$, from which the next searching step will be performed (\textit{line 3}, Fig.~\ref{fig:algorithm}(a)). 
The second step searches the design space by creating a set of data facts (denoted as $F_i$), that is logically relevant to $f_i$ (\textit{line 4}, Fig.~\ref{fig:algorithm}(b)).
The third step finds the best searching direction $f_i \rightarrow f^*, f^* \in F_i$ with the largest reward $\Delta^*$ through a simulation process (\textit{lines 5 - 11}, Fig.~\ref{fig:algorithm}(c)). This process simulates the cases in which each $f \in F_i$ is expanded in a simulation tree $\mathcal{T}_s$ rooted at $f_i$ to help the algorithm explore the space a few steps further, so that the different searching directions can be estimated in advance. The simulation runs within a time limit to ensure the efficiency of the algorithm (\textbf{R2}).
In the last step, the tree $\mathcal{T}$ is updated via a back-propagation process, in which the weights of the relevant nodes are updated based on $\Delta^*$ and $f^*$ is formally added into $\mathcal{T}$ as a child of $f_i$ (\textit{line 11}, Fig.~\ref{fig:algorithm}(d)). Finally the path with the highest reward in $\mathcal{T}$, $\mathcal{P}^*$, is identified as the best story generated at the current iteration (\textit{line 12}). The algorithm stops when the goal $\mathcal{G}$ is fulfilled.


\setlength{\textfloatsep}{0pt}
\begin{algorithm}[tb]
\label{alg:generation}
\SetAlgoLined
\SetKwInOut{Input}{Input}
\SetKwInOut{Output}{Output}
\Input{$\mathcal{D}, \mathcal{G}$}
\Output{$\mathcal{S} = \{f_1,r_1 \cdots, f_{n-1}, r_{n-1}, f_n\}$}
$f_0$ $\leftarrow$ Initialize($\mathcal{D}$);
$\mathcal{T}$ $\leftarrow$ \{$f_0$\};
$\mathcal{S}$ $\leftarrow$ \{\}\;
\While{$\mathcal{G}$ is not fulfilled}{
    \tcc{1.\selection}
    $f_i \leftarrow select(\mathcal{T})$\; 
    
    \tcc{2.\expansion}
    $F_i$ $\leftarrow$ Expand($f_i$)\;
    
    \tcc{3.\simulation}
    $\mathcal{T}_s \leftarrow \{f_i\}$;
    $F \leftarrow F_i$;
    $f_p \leftarrow f_i$\;
    \While{within time limitation}{
    
        \tcp{Calculate the reward of each node in $F$ in context of $\mathcal{T}_s$ and find the node $f \in F$ with the highest reward $\Delta$. The design space will be explored in direction of $f_p \rightarrow f$ in the simulation process.}
        $f, \Delta \leftarrow$ Reward($\mathcal{T}_s, F$)\;
        
        \tcp{Add $f$ in the simulation tree $\mathcal{T}_s$ as a child of $f_p$ and update reward of the relevant nodes in $\mathcal{T}_s$ based on $\Delta$. After that, find the node $f^* \in F_i$ with the highest reward in $\mathcal{T}_s$, where $f_i \rightarrow f^*$ determines the best searching direction found so far.}
        $f^*,\Delta^* \leftarrow$ BackPropagation($\mathcal{T}_s, f, \Delta$)\;
        
        \tcp{Select the next node $f_p$ and expand it in the simulation tree for a further exploration}
        $f_p \leftarrow select(\mathcal{T}_s)$;
        $F$ $\leftarrow$ Expand($f_p$)\;
    }
    \tcc{4.\backprop}
    BackPropagation($\mathcal{T},f^*, \Delta^*$)\;
    
    $\mathcal{S} \leftarrow \mathcal{P}^* = \{f_1, r_1, \cdots, f_{n-1}, r_{n-1}, f_n\}$\;
}
\Return $\mathcal{S}$\;
\caption{Logic-Oriented Monte Carlo Tree Search}
\end{algorithm}

\paragraph{\bf Logic-Oriented Node Expansion} Expanding a selected node $f_i$ in the search tree $\mathcal{T}$ to elaborate the story design space is a critical step in the aforementioned searching algorithm. The expansion should generate a set of nodes that is logically relevant to $f_i$ to gradually generate a meaningful data story through the searching process. To this end, we investigated how a set of commonly used coherence relations (denoted as $\mathcal{R}$)~\cite{wellner2006classification, wolf2005representing} was used in data stories during our preliminary survey. As a result, Table~\ref{tab:logic} summarizes the likelihood, $P(r_i|f_i)$, of each relation $r_i$ occurring after a fact $f_i$ regarding to their fact types, which guides the node expansion process. In particular, during the expansion, we create a set of data facts regarding each coherence relation. The proportion of the newly generated facts is given by $P(r_i|f_i)$, and each new fact, $f_{i+1}$, is generated by the following rules:
\begin{itemize}[leftmargin=10pt,topsep=2pt]
\itemsep -.5mm
    \item {\textit{\textbf{Similarity}}} indicates two succeeding facts are logically parallel to each other. Therefore, $f_{i+1}$ can be generated by a variety of methods, such as modifying the measure / breakdown / focus field without changing the subspace. 
    \item {\textit{\textbf{Temporal}}} relation communicates the ordering in time of events or states. In this case, we generate $f_{i+1}$ by changing the value of the temporal filter in $f_{i}$'s subspace to a succeeding time.
    \item {\textit{\textbf{Contrast}}} indicates a contradiction between two facts. For simplicity, we only check the contradictions in two types of facts, i.e., trend and association. $f_{i+1}$ is generated by modifying the subspace of $f_i$ to form a data contradiction in measures. For example, the sales trends of a product increases, but that of another product decreases. The sales number of a product is positively associated with its price, but the association is negative in case of another product. In these examples, the subspace is determined by different products.
    \item {\textit{\textbf{Cause-Effect}}} indicates the later event is caused by the former one. \rv{In multidimensional data, a causal relation can be determined between dimensions based on the data distribution}. In this way, $f_{i+1}$ can be generated by changing the measure field $m_i$ of $f_i$ to another numerical field in the spreadsheet that is most likely caused by $m_i$ in accordance with causal analysis~\cite{schaechtle2013multi}.
    \item {\textit{\textbf{Elaboration}}} indicates a relation in which a latter fact $f_{i+1}$ adds more details to the previous one $f_i$. In this way, we create $f_{i+1}$ by shrinking the scope of $f_i$'s subspace via adding more constraints (i.e., filters) or setting a focus to ``zoom" $f_i$ into a more specific scope. 
    \item {\textit{\textbf{Generalization}}} indicates $f_{i+1}$ is an abstraction of the previous $f_i$, which is in opposite to elaboration. Therefore, we create $f_{i+1}$ by enlarging the scope of $f_i$'s subspace via removing constraints.
\end{itemize}

\paragraph{\bf Reward Function} We propose a reward function that estimates the quality of each generated story $\mathcal{S}$ via three criteria, i.e., diversity $D(\mathcal{S})$, logicality $L(\mathcal{S})$, and integrity (i.e, data coverage) $C(\mathcal{S})$ based on the story's information entropy $H(\mathcal{S})$:
\begin{equation}
reward(\mathcal{S}) = \{\gamma_1 \cdot D(\mathcal{S}) + \gamma_2 \cdot L(\mathcal{S}) + \gamma_3 \cdot C(\mathcal{S})\} \cdot H(\mathcal{S})
\label{eq:reward}
\end{equation}
where $\gamma_i \in [0,1], \sum_i \gamma_i = 1$ are the weighting parameters given by users to balance different criteria. All the criteria are also normalized to $[0,1]$. $H(\mathcal{S})$ is the story's information entropy that indicates the expected self-information of the data facts in the story, which is used as the basis of the reward and formally defined as follows:
\begin{equation}
    H(\mathcal{S})= \sum_{i=1}^{n} { P(f_i) \cdot I_s(f_i) } = -\sum_{i=1}^{m} P(f_i) \cdot S(f_i) \cdot \log _{2}\left(P(f_i)\right)
\end{equation}
where, $I_s(f_i)$ is the fact's importance score defined in Formula (\ref{eq:importance}). The definition of each story estimation criterion is described as follows:

\begin{itemize}[leftmargin= 10pt]
\itemsep -.5mm
\item \textbf{\textit{Diversity}} estimates the variance of the fact types in $\mathcal{S}$. Rich fact types will make the story vivid and attractive. Diversity is given by two terms as follows:
\begin{equation}
D(\mathcal{S}) = \frac{n}{\min(|\mathcal{S}|,10)} \cdot \frac{-\sum_{i=1}^{n} {p}_{i} \cdot \ln \left({p}_{i}\right)}{\ln (n)} 
\end{equation}
where $n$ indicates the total number of fact types used in $\mathcal{S}$ whose maximum value is known as 10; $p_i$ is the proportion of the $i$-th fact type in the story. When $D(\mathcal{S})$ is maximized, the first term encourages to use more fact types in a story, and the second term, a normalized Shannon diversity index~\cite{ramezani2012note}, ensures that different fact types can be evenly used in the story.

\item \textbf{\textit{Logicality}} estimates the logical coherence of a story. A higher logicality score indicates the story is more coherent and easier to follow. Logicality is defined by the averaged likelihood of each coherent relation $r_i$ occurred after each fact $f_i$ in the story: 

\begin{equation}
L(S) = \frac{1}{n-1}\sum_{r_i,f_i \in S}P\left(r_{i} | f_{i}\right)
\end{equation}

\item \textbf{\textit{Integrity}} is the data coverage rate of $\mathcal{S}$. A larger integrity, indicates the story more comprehensively represents the input data. Integrity is defined as:
\begin{equation}
C(\mathcal{S}) = \frac{count(\bigcup\limits_{i=0}^{n-1} f_{i})}{\mathcal{N}}
\end{equation}
where the molecule is the total number of data items in the spreadsheet used in the story, and $\mathcal{N}$ is the total number of data items.
\end{itemize}
\section{Data Story Editor}
\label{sec:editor}

\begin{figure}[t]
  \centering
  \includegraphics[width=\linewidth]{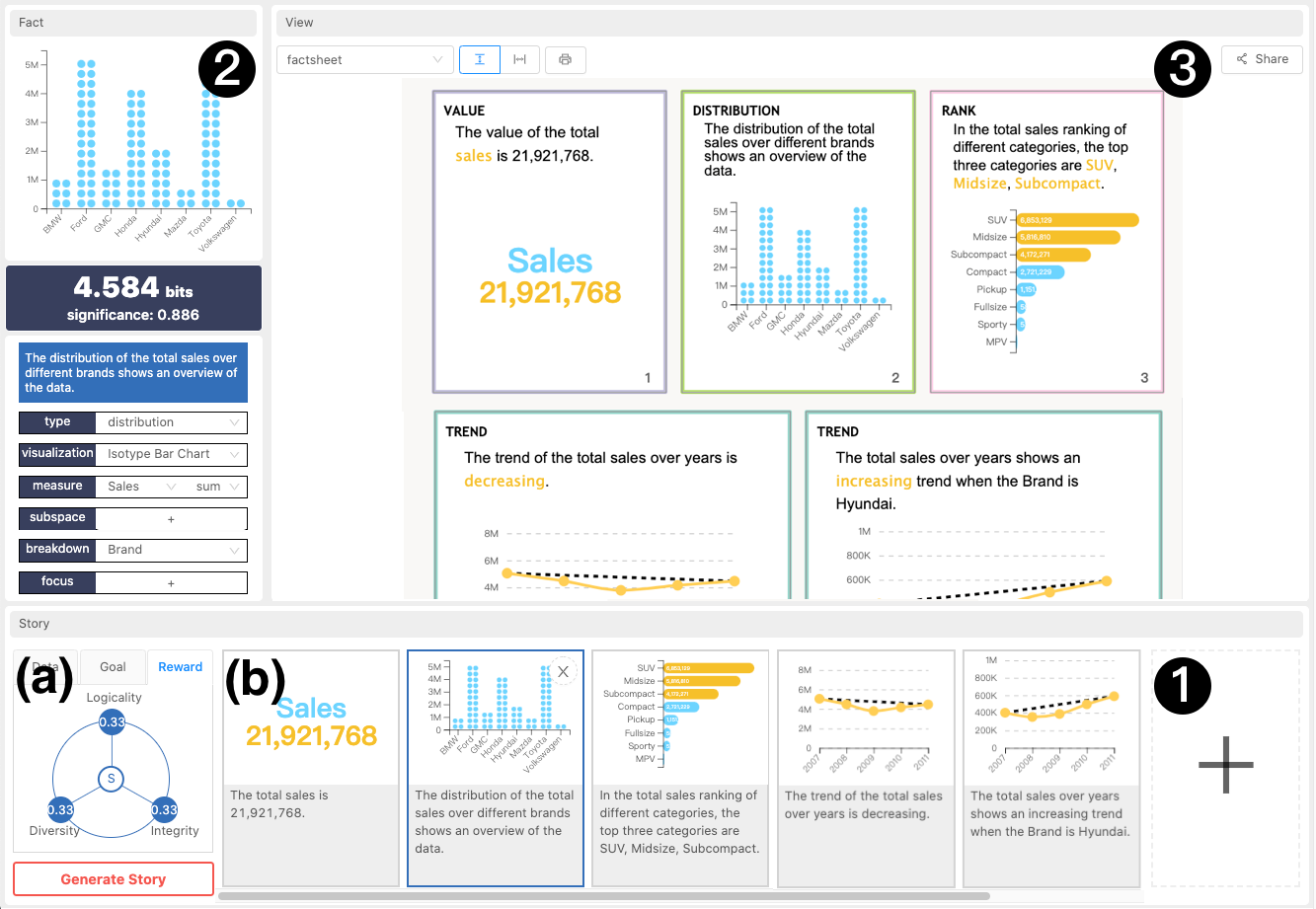}
  \vspace{-1.5em}
  \caption{The story editor of \name system consists of three major views: (1) the storyline view for story configuration, generation, and storyline editing, (2) the fact view for fact editing, and (3) the story visualization view for the visual data story preview and sharing \rv{(\factsheet mode)}. 
    } 
  \label{fig:ui}
  \vspace{-1em}
\end{figure}
\setlength{\textfloatsep}{20pt}

In this section, we introduce the design of the story editor and the methods used for visualizing a data story.

\subsection{User Interface}
\begin{figure*}[htb]
  \centering
  \includegraphics[width=0.86\linewidth]{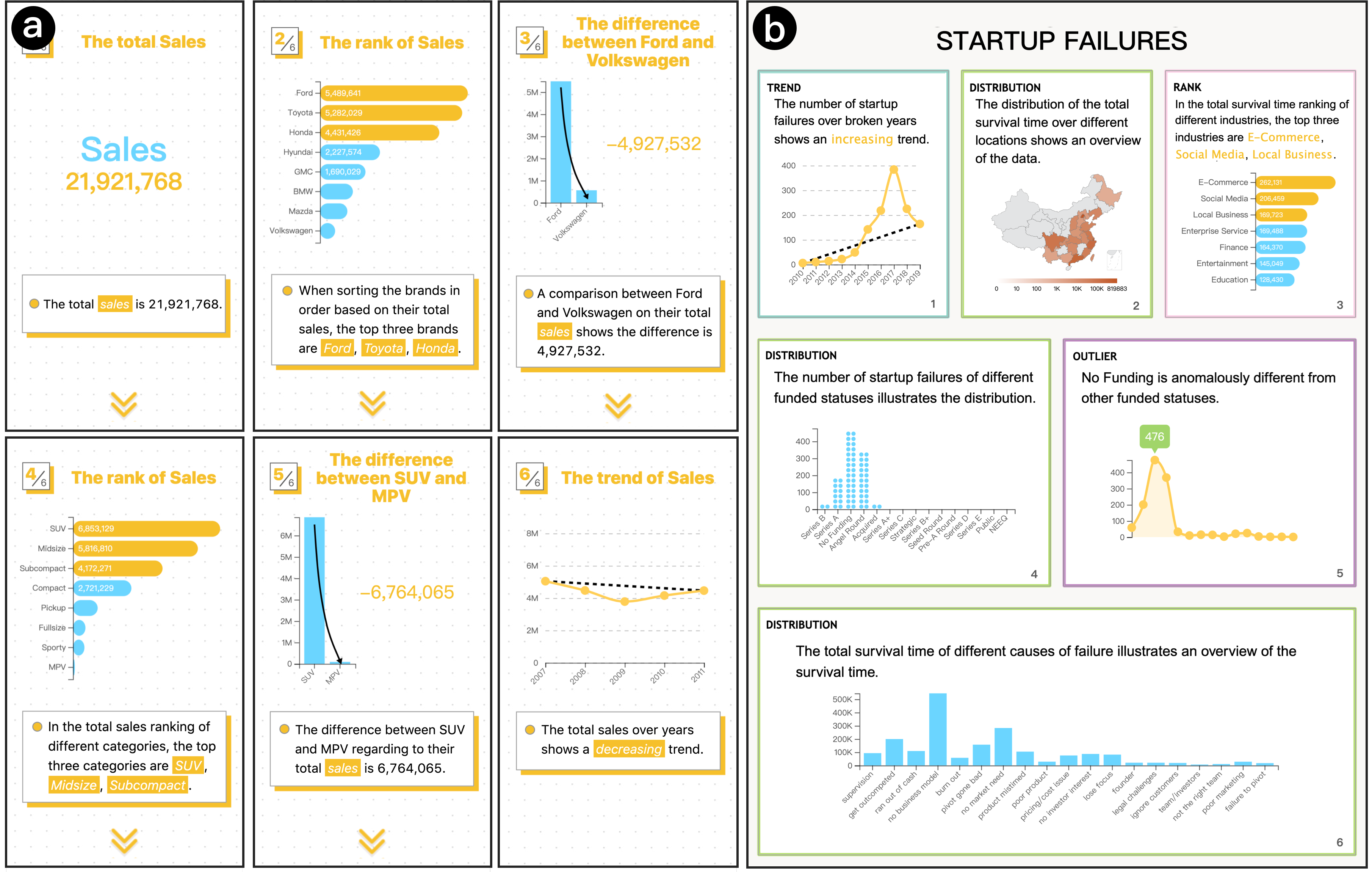}
  \vspace{-0.5em}
  \caption{Two data story examples generated by \name: (a) a story about car selling records around economic crisis in 2008 shown in the \swiper mode and (b) a story about startup failures after the tide of ``new economics" in China shown in the \factsheet mode.} 
  \label{fig:casestudy}
  \vspace{-1em}
\end{figure*}

The story editor, as shown in Fig.~\ref{fig:ui}, consists of three major views: the storyline view (Fig.~\ref{fig:ui}-1), fact view (Fig.~\ref{fig:ui}-2), and story visualization view (Fig.~\ref{fig:ui}-3). In the storyline view, a user can upload a spreadsheet, set the story generation goal, and adjust the reward function in a group of configuration panels (Fig.~\ref{fig:ui}-1(a)). The generated data facts are shown in a row (Fig.~\ref{fig:ui}-1(b)), in which a user can remove a fact or change the generated narrative order based on his/her own preferences. Each fact is visualized by a chart and captioned by a generated text description (\textbf{R3}). When a fact is selected, the data details on each of its fields, importance scores, and visual and textual representations, will be shown in the fact view (\textbf{R4}). \rv{The generated data story can be visualized in the story visualization view through three visualization modes: (1) \storyline mode (Fig.~\ref{fig:teaser}), (2) \swiper mode (Fig.~\ref{fig:casestudy}(a)), and (3) \factsheet mode (Fig.~\ref{fig:casestudy}(b)). These modes are respectively designed for representing the story on laptops/tablets, smartphones, and printouts to facilitate a flexible story communication and sharing (\textbf{R5}). A user can easily switch between different 
modes in the story visualization view 
via a drop down menu.}


\subsection{Visualizing a Data Story}
A data story generated by the engine is visualized as a sequence of charts through two steps: \textit{showing a data fact} and \textit{showing a story}. The first step maps a data fact to a chart while the second step organizes a sequence of charts in an expressive layout as a story.

\paragraph{\bf Showing a Data Fact} Benefiting from the simple and clear definition of each fact type introduced in Section~\ref{sec:fact}, \name is able to directly convert a data fact into a captioned chart that incorporates both the visual and textual representations. Specifically, the caption is generated based on the syntax defined in Table~\ref{tab:fact}, and the fact is automatically visualized in two steps by following a rule-based approach. In particular, the system first selects the most frequently used chart regarding the fact type in Table~\ref{tab:charts}. After that, it selects a subset of data from the input spreadsheet regarding the filters given by the \subspace field and then maps the \breakdown field(s) to the categorical channel(s) and the \measure field(s) to the numerical channel(s) in the chart. Finally, the data values indicated by the \focus field are highlighted in the chart. Fig.~\ref{fig:ui}(2) illustrates an example of showing a difference fact in a captioned bar chart.


\paragraph{\bf Showing a Story} The story visualization view provides a variety of visualization modes to represent the generated data story for different application scenarios. In particular, a summarization is first provided to give a textual briefing of the story to help users obtain data insights at a glance. This step shows the data coverage rate, the total number of data facts in the story, and the generated textual narrative of the story. The \storyline and \swiper visualization modes are respectively designed to facilitate an efficient story exploration on tablets and smartphones. \rv{In particular, the captioned charts showing data facts in the story are horizontally aligned in a row to represent the narrative order in the \storyline mode} and are shown one at a time in the \swiper mode. A user can swipe the touch screen to explore the story through these representations. Finally, the \factsheet mode is designed to show the story in the form of a poster that could be easily printed out.


\section{Evaluation}
We evaluate \name system via (1) three examples to showcase the quality of the generated story, (2) two controlled experiments to estimate the usefulness and quality of a generated logic, and (3) domain expert interviews to estimate the usability of the system.

\subsection{Example Stories}
We demonstrate three visual data stories generated by \name respectively based on three real-world datasets as shown in Fig.~\ref{fig:teaser} and Fig.~\ref{fig:casestudy}, which are described as follows:

Fig.~\ref{fig:teaser} shows a story generated based on a \Covid dataset (903 rows, 5 columns). The data record the recent numbers of daily infections, deaths, and healings of COVID-19 in China from March 1st to March 21st. The generated story illustrates the daily mortality in China decreased in March ({\it Fact 1}), and the largest number was 42 occurred on March 2nd ({\it Fact 2}). Hubei was the most affected province ({\it Fact 3}). The total death in Hubei accounted for 97.4\% of that in China ({\it Fact 4}), which was 423 ({\it Fact 5}). A large number of patients recovered in March ({\it Fact 6}), showing the improving situation in China.

Fig.~\ref{fig:casestudy}(a) shows a story generated based on a \CarSales dataset (275 rows, 4 columns), including the sales records of different automobile brands around the financial crisis in 2008. The story shows that in 2007-2011, 21,921,768 cars were sold in total ({\it Fact 1}). The top three sellers were Ford, Toyota, and Honda ({\it Fact 2}). The difference was huge when comparing the best and worst sellers ({\it Fact 3}). Specifically, SUV was the best selling model ({\it Fact 4}) which sold 6,764,065 more than MPV ({\it Fact 5}). Generally, the sales records decreased during the final financial crisis ({\it Fact 6}).

Fig.~\ref{fig:casestudy}(b) presents a dataset about \Startup . The data (1234 rows, 6 columns) record a set of companies closed during or after the raising tide of ``new economics" in China from 2010 to 2019. Each startup company is described from six criteria, including its broken year, location, industry, funded status, survival time, and the main cause of failure. The story shows that numerous companies were closed in recent years ({\it Fact 1}), and most of them were located in Eastern China ({\it Fact 2}). The most dangerous fields were e-commerce, social media, and local business ({\it Fact 3}). Most companies closed in these fields were still in the early stages before the series A+ round ({\it Fact 4}), and some even closed without receiving any investment ({\it Fact 5}). Regarding the reasons, ``no business model” and ``no market need” were the most frequently occurring problems in these startup companies ({\it Fact 6}).


\subsection{Evaluation of the Generated Logic}
We verified the usefulness and the quality of the generated logic in a story via two controlled experiments.

\paragraph{\bf Experiment I: Usefulness}
We first estimated whether the logic generated by \name helps with the understanding of a data story. To this end, we compared our generation results with the factsheets generated by DataShot~\cite{wang2019datashot}, in which a set of selected data facts was randomly organized. 

{\underline{\textit{Data.}}}
\rv{We collected the same datasets illustrated in Fig. 4 (C, D) in DataShot, including CarSales and SummerOlympics. Based on these two datasets, we generated two factsheets using Calliope automatically and picked two cases from the DataShot paper directly as the baseline. To ensure a more fair study setting, we made sure these generated stories from different systems contained similar data facts, and we also revised the design of each factsheet with the same style.}

{\underline{\textit{Procedure and Tasks.}}}
\rv{We recruited 16 college students (12 females) ages 22 - 26 years old (M=23.94, SD=1.43) as participants. Each participant was presented with two factsheets of different topic from our data, one by Calliope and one by DataShot.
We counterbalanced the presentation order of the factsheets for a fair comparison.
The participants were asked to read the factsheets carefully and compare two factsheets from five specific aspects including logicality, comprehension, memorability, engagement, and dissemination. The experiment lasted approximately 40 minutes for each participant.}

{\underline{\textit{Results.}}}
\rv{In terms of \textit{Logicality} and \textit{Memorability}, Calliope received more positive feedback than DataShot. One participant commented that ``\textit{I can smoothly follow it's (Calliope) logic from whole to part, as it first introduces the overall information about Olympic golds and then zooms in on specific sports and countries.}" Another participant said, ``\textit{it's much easier to remember the story generated by Calliope, as the annual car sales in different brands is presented step by step in a proper order}". 
Regarding \textit{Comprehension}, \textit{Engagement}, and \textit{Dissemination}, Calliope performs comparably to Datashot. One participant said, ``\textit{I enjoy the 
simple and beautiful visualization of both factsheets and would love to share them on social media if the data is relevant.}"} 

\paragraph{\bf Experiment II : Quality} 
The second experiment was designed to evaluate the quality of the generated logic. To this end, we objectively estimated consistency of the logic orders respectively given by users and generated by \name based on the same set of data facts.

\underline{\textit{Procedure and Tasks.}} We first shuffled the order of a set of data facts in a visual data story generated by \name and then asked a group of users to restore the logic order by reading the chart and description of each fact. Finally, we checked the consistency between the human-generated logic orders and those produced by \name based on Kendall's $\tau_b$ correlation~\cite{kendall1945treatment}. 
\rv{This measure was introduced to estimate the consistency of the element orders between two sequences, whose value lies in $[-1, 1]$ with ``-1" indicating a completely reverse order and ``1" indicating the orders are identical}. To ensure a fair and comprehensive comparison, we generated 12 data stories based on the aforementioned three datasets, four stories per dataset. Each story contained six data facts whose orders were shuffled for the experiment. 

A group of 20 participants (17 female) aged 22-30 years old ($M=26, SD=2.63$) were recruited for Experiment II. All of the participants reported that they have fundamental knowledge about data visualization or experience in data-oriented storytelling. The experiments started by a brief introduction about the data, and the participants were asked to reorder the data facts of all the aforementioned 12 shuffled stories via an interactive user interface. 
We also encouraged the participants to fully explore the data and try their best to understand the data insights represented by each data fact.

{\underline{\textit{Results.}}}
\rv{We calculated the average Kendall's $\tau_b$ value on each dataset, and the result showed that the logic orders generated by \name were consistent with those generated by our participants (\CarSales: $\mu=0.487, \sigma=0.29$; \Covid: $\mu=0.648, \sigma=0.327$; \Startup: $\mu=0.63, \sigma=0.295$). 
We also leveraged the sequences generated by humans as a ground truth and calculated Kendall's $\tau_b$ values between the random results and human generated results as the baseline. As a result, the baseline is 0.015, which indicates these two orders are relatively irrelevant (as the value is close to 0). 
By comparing the $\tau_b$ values, we found that the logic orders generated by Calliope are much more consistent to humans than the baseline.}

\begin{figure}[b]
  \centering
  \includegraphics[width=0.9\linewidth]{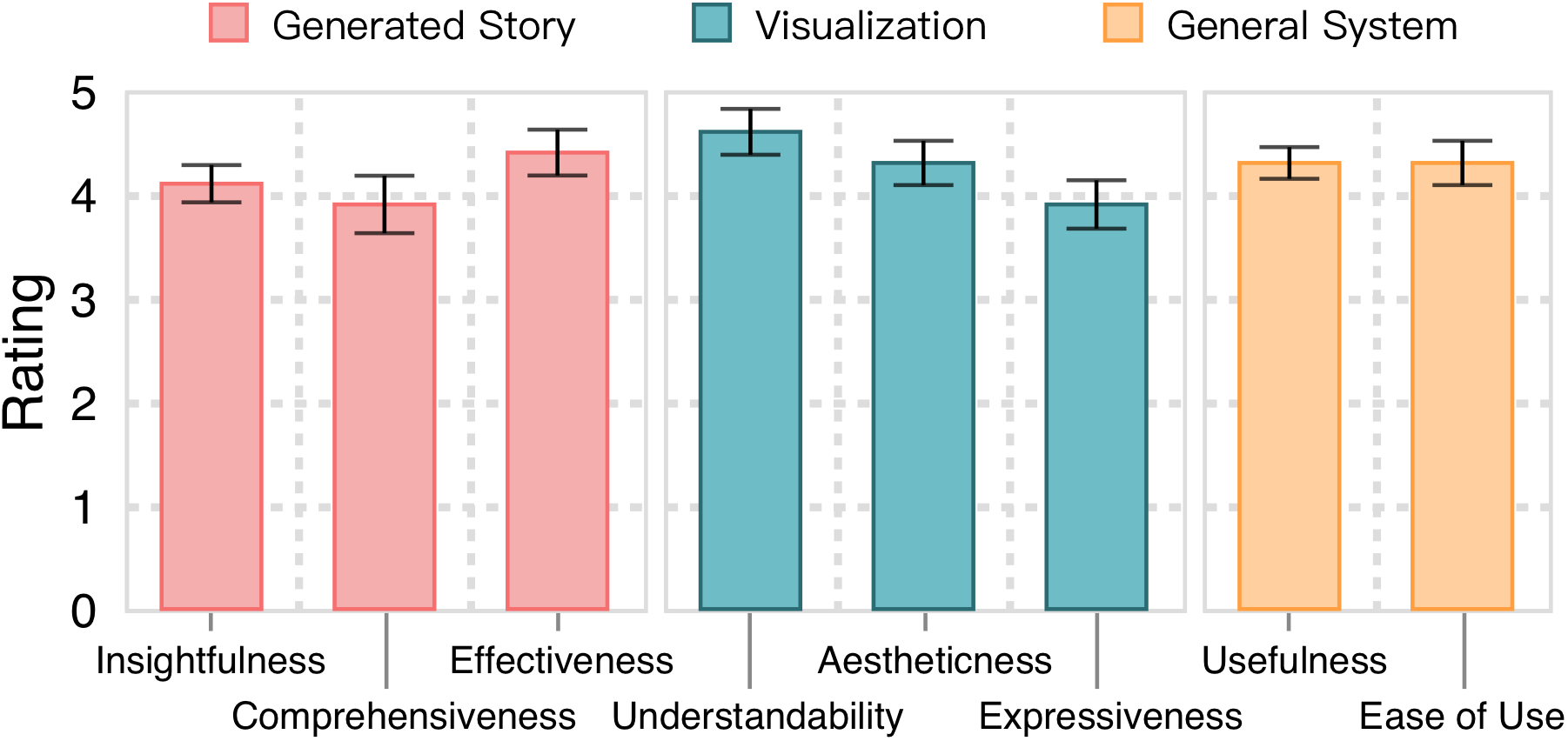}
  \vspace{-0.5em}
  \caption{The ratings of generated story, visualization design, and system from different criteria based on a 5-point Likert scale given by 10 expert users, where 5 is the best and 1 is the worst.}
  \label{fig:questionare}
  \vspace{-0.5em}
\end{figure}

\subsection{Expert Interview}
To further evaluate the usability of \name system, a series of interviews with three groups of expert users from different areas were performed. The first group included four data journalists (denoted as J1-J4) from different news medias in China. They had over 3.5 years' working experiences on average, and were very familiar with the data-oriented storytelling and had technical skills for creating a visual data story. The second group comprised three data analysts (denoted as D1-D3) from an international IT consultant company. Their major job was to analyze customers' data and write analysis reports. BI tools such as Tableau and PowerBI were frequently used in their daily work. The third group consisted of three senior visualization researchers (denoted as V1-V3), all of whom had experiences on publishing papers in major visualization conferences such as IEEE VIS and EuroVis. 

\paragraph{\bf Procedure} The interviews were performed via an online meeting system. Each interview started by a 10-minutes introduction about the system. After that, the experts were encouraged to use the online \name system on their own. After fully explored the functions of the system, the experts were asked to generate a data story based on one of the three datasets introduced in Experiment II. To arouse their interests, we let the journalists explore the \Covid dataset as it is a recent and important news topic. We let the data analysts use the \CarSales data given its similarity to the business data analyzed in their work. We let the visualization researchers explore the \Startup data as it is the most complex one. They were also encouraged to edit the generated story and share their findings with us. The experts were asked to finish a questionnaire, followed by an interview after they fully explored the functionalities of \name system. The whole process lasted for about one hour and the interviews were recorded for later analysis.

\paragraph{\bf Results} 
Fig.~\ref{fig:questionare} shows that questionnaire results, \name was rated relatively high in terms of the generated story, visualization, and system.
A large number of positive comments were recorded during the interview. We summarized the interview results due to the page limitation.

\textit{\underline{Data Story.}} 
All the experts agreed that the generated data story was able to express useful data insights. The visualization researcher V2 mentioned that ``\textit{the data facts in the story are quite clear and are well-organized}". J1 commented that ``\textit{the story starts from an overview, followed by a series of data details, ..., It's the way we frequently adopted when writing a news story. It's amazing that now it can be automatically generated}". J3 also noted that \textit{``the logical order [of the story] can help readers get into the points"}. D1 observed that the ordered data facts can help with the efficient exploration of the data,  \textit{``[Automatically] showing the appropriate dimensions and measures in a sequence of visualizations can definitely guide the data exploration, ..., it's helpful when you have no idea about how to get started"}.

\textit{\underline{Visualization.}}
In terms of visualization, all the experts were satisfied with the design of three visualization modes in \name, \textit{``it's a very nice and thoughtful design"} (J1-J4). Especially, they felt that the \storyline mode provides a good overview (J1-J3, D1, D3, V1, V2). All the experts believed that the \swiper mode was neat and helpful when viewing the story on a smartphone while the \factsheet mode supports easy printing of a story. 
All the journalists felt that editing was an important function, \textit{``it (editing function) allows us to create a high-quality story quickly based on the generation results"} (J1). They also felt generating stories by interactively changing the reward was \textit{``interesting"} and \textit{``inspiring"}. J2 suggested that \textit{``we usually write stories from different perspectives, and it can facilitate my ideation process"}.

\textit{\underline{System.}}
Most experts (J1-J3, D1, D3, V2, V3) mentioned that the system was useful for users who are not skilled at data analysis or visualization. The data journalists J1-J4 especially appreciated the efficient story generation and editing function of \name. J4 said, \textit{``with this tool I can quickly create a story by first generating a draft and then revising it accordingly"}. The data analysts felt the system is powerful to help them efficiently preview an unknown data, \textit{``with this tool, I can quickly find where to start when getting a [new] spreadsheet"} (D1). The visualization researchers believed that our system lowers the barriers of creating a data story. V3 said, ``when compared with other data story authoring tools, this system is much more smart as it requires limited knowledge about data analysis and visualization design". 

\section{\bf Limitations and Future Work}
\rv{Despite the above positive results from the evaluation, we would also like to summarize and discuss several limitations that was found during the design and implementation process and mentioned by our expert users during the interview. 
We hope highlight these limitations will help point out several potential future research directions and inspire new studies by following our work.
}

\rv{\underline{\textit{Supporting a Better Textual Narrative.}} During the interview, many data journalists (J1,J2,J4) felt the generated captions were too rigid to be used especially in a data news. More diverse and insightful descriptions were desired. Moreover, the current results also contained some grammar errors, which also need to be addressed (J1,2 D1, V2,3)). However, all of them acknowledged that the current results, although unsatisfactory, were still useful for a rapid preview and briefing of the input data. In the future, it is necessary to leverage more advanced techniques introduced in the field of nature language processing to generate text narratives in the data story in a higher quality.}

\underline{\textit{Understanding Data Semantics.}} After using \name, although impressed, some experts (J1-J3, V2) expect a more intelligent tool that can even understand the semantics of the data to better generate the story content and logic. \rv{We acknowledge this is a key limitation of the current system and understanding the underlying semantics of the data is critical for generating a meaningful and insightful data story. This is a promising research direction that is worth a further exploration. To address the issue, one could leverage or develop more advanced AI techniques or could also introduce sophisticated interactive feedback mechanism to keep user in the generation loop and leverage human intelligence to 
steer data quality~\cite{liu2018steering} and guide the underlying generation algorithm/model to better understand data semantics.}

\underline{\textit{Enriching Visualization.}} Several experts (J3, D1, D2) would like to have a slides mode and dashboard mode to support more application cases. J1 and V3 also pointed out that some of the current generated visual encoding are notably simple, and a chart should encode more information at a time. For example, when showing a line chart, the size of point could also be used to encode data, and a stack bar chart could be used to show an additional categorical field in the data. \rv{In addition, \name cannot deal with hierarchical or relational datasets, which are also desired functions (V1,V2). Providing more advanced visualization representations for a story is also a valuable future work.}

\underline{\textit{Performance Issues.}}
The current system design and implementation have some performance bottlenecks that worth a future study.  \rv{In particular, the calculation of data fact significance, i.e., $S(f_i)$ in Formula (\ref{eq:importance}) consists of statistical computations, which is usually slow and thus limits the number of facts that can be explored in each searching iteration, thus affecting the generation quality.}
\section{Conclusion}

We have presented \name, a novel system designed for automatically generating visual data stories from a spreadsheet. The system incorporates a novel logic-oriented Monte Carlo tree search algorithm to create a data story by gradually generating a sequence of data facts in a logical order while exploring the data space. The importance of a fact is measured by its information quantity and its statistical significance. Each fact in the story is visualized in a chart with an automatically generated caption. A story editor is introduced in the system to facilitate the easy and efficient editing of a generated story. The proposed technique was evaluated via three example cases, two controlled experiments, and a series of interviews with 10 expert users from different domains. The evaluation showed the power of \name system and revealed several limitations of the current system, which will be addressed in the future.

\acknowledgments{
Nan Cao is the corresponding author.
This work was supported by NSFC 62072338 and NSF Shanghai 20ZR1461500.
}

\bibliographystyle{abbrv-doi}

\bibliography{main}
\end{document}